\documentclass[%
 reprint,
 superscriptaddress,
frontmatterverbose, 
showpacs,preprintnumbers,
 amsmath,amssymb,
 aps,
prl,
floatfix,
]{revtex4-1}

\usepackage{graphicx}
\usepackage{dcolumn}
\usepackage{bm}
\usepackage{url}
\usepackage[utf8]{inputenc}
\usepackage[caption=false]{subfig}
\usepackage{float}

\begin{document}


\title{Magneto-structural correlations in a systematically disordered B2 lattice}

\author{Jonathan Ehrler}
  \email{j.ehrler@hzdr.de}
  \affiliation{Helmholtz-Zentrum Dresden-Rossendorf, Bautzner Landstrasse 400, 01328 Dresden, Germany}
  \affiliation{Technische Universität Dresden, Helmholtzstrasse 10, 01069 Dresden, Germany}
\author{Biplab Sanyal}
  \affiliation{Department of Physics and Astronomy, Uppsala University, Box-516, 75120 Uppsala, Sweden}
\author{Jörg Grenzer}
  \affiliation{Helmholtz-Zentrum Dresden-Rossendorf, Bautzner Landstrasse 400, 01328 Dresden, Germany}  
\author{Shengqiang Zhou}
  \affiliation{Helmholtz-Zentrum Dresden-Rossendorf, Bautzner Landstrasse 400, 01328 Dresden, Germany}
\author{Roman Böttger}
  \affiliation{Helmholtz-Zentrum Dresden-Rossendorf, Bautzner Landstrasse 400, 01328 Dresden, Germany}
  \author{Benedikt Eggert}
  \affiliation{Faculty of Physics and Center for Nanointegration Duisburg-Essen (CENIDE), University of Duisburg-Essen, Lotharstraße 1, D-47048 Duisburg, Germany}
\author{Heiko Wende}
  \affiliation{Faculty of Physics and Center for Nanointegration Duisburg-Essen (CENIDE), University of Duisburg-Essen, Lotharstraße 1, D-47048 Duisburg, Germany}
\author{Jürgen Lindner}
  \affiliation{Helmholtz-Zentrum Dresden-Rossendorf, Bautzner Landstrasse 400, 01328 Dresden, Germany}
\author{Jürgen Fassbender}
  \affiliation{Helmholtz-Zentrum Dresden-Rossendorf, Bautzner Landstrasse 400, 01328 Dresden, Germany}
\author{Christoph Leyens}
  \affiliation{Technische Universität Dresden, Helmholtzstrasse 10, 01069 Dresden, Germany}
\author{Kay Potzger}
  \affiliation{Helmholtz-Zentrum Dresden-Rossendorf, Bautzner Landstrasse 400, 01328 Dresden, Germany}
\author{Rantej Bali}
  \email{r.bali@hzdr.de}
  \affiliation{Helmholtz-Zentrum Dresden-Rossendorf, Bautzner Landstrasse 400, 01328 Dresden, Germany}

\date{\today}

\begin{abstract}

Ferromagnetism in certain B2 ordered alloys such as Fe\textsubscript{60}Al\textsubscript{40} can be switched on, and tuned, \textit{via} antisite disordering of the atomic arrangement. The disordering is accompanied by a $\sim$1\,\% increase in the lattice parameter. Here we performed a systematic disordering of B2 Fe\textsubscript{60}Al\textsubscript{40} thin films, and obtained correlations between the order parameter\,($S$), lattice parameter\,($a_0$), and the induced saturation magnetization\,(M\textsubscript{s}). As the lattice is gradually disordered, a critical point occurs at \hbox{1-$S$}=0.6 and $a_0$=2.91\,\AA, where a sharp increase of the M\textsubscript{s} is observed. DFT calculations suggest that below the critical point the system magnetically behaves as it would still be fully ordered, whereas above, it is largely the increase of $a_0$ in the disordered state that determines the M\textsubscript{s}. The insights obtained here can be useful for achieving tailored magnetic properties in alloys through disordering.

\end{abstract}

\maketitle

Controlled disordering of the crystal lattice can unlock potential to tune the properties of magnetic materials. Intrinsic material properties such as the saturation magnetization (M\textsubscript{s}) can be highly sensitive to the ordering of atoms in alloy lattices \cite{Zeng2006,Avar2014,Hernando1998,Amils1998,Heidarian2015,Sanchez1996,Krause2000}, manifesting a wide range for M\textsubscript{s}-tuning. However, a precise understanding of the mechanism of increasing M\textsubscript{s} with decreasing structural order has been elusive, since varying the ordering also causes changes of other structural properties. This makes it difficult to experimentally associate the changes of the magnetic properties to various changes in structural properties.

The understanding of disorder-induced effects can be approached in prototype alloys that respond sensitively to disordering. Examples of this behavior are certain binary alloys which order in the B2 structure and transform from para- or antiferromagnets to ferromagnets \textit{via} small atomic re-arrangements \cite{Zeng2006,Avar2014,Hernando1998,Amils1998,Heidarian2015,Sanchez1996,Krause2000}. 

In this study, we selected a Fe\textsubscript{60}Al\textsubscript{40} alloy that can form an ordered B2 phase (referred to as B2 Fe\textsubscript{60}Al\textsubscript{40}) which is paramagnetic (PM) and transitions to ferromagnetism as the structure is chemically disordered \cite{Zeng2006,Avar2014,Hernando1998,Amils1998,Bradley1932,Menendez2008,Beck1971,Fassbender2008,Bali2014,Tahir2015,Roeder2015,Liedke2015,Ehrler2018}. The B2 phase consists of a superstructure of two interpenetrating simple cubic lattices; one each lattice for the Fe and Al atoms respectively \cite{Kuentzler1983}. Disordering implies a rearrangement to form antisite defects, whereby an Fe(Al) lattice site is replaced with an Al(Fe) atom (Figure 1a). Randomization of the Fe and Al site occupancies \textit{via} disordering leads to the A2 structure. This B2 to A2 transition results in an increase in the number of the Fe-Fe nearest neighbors, from an average of 2.7 in B2 to 4.8 in the A2; this increase has often been a first-approximate explanation for the onset of ferromagnetism \cite{Beck1971,Zamora2009,Huffman1967,Takahashi1990}. However, the disordering process is accompanied by a lattice expansion of $\sim$1\,\% \cite{Hernando1998,Amils1998,Nogues2006,Fujii1999,Apinaniz2001,Apinaniz2003,Apinaniz2004,Smirnov2005,Rodriguez2012,Plazaola2012}.
 
To investigate the role of strain, recent studies have applied mechanical deformation to induce disorder \cite{Hernando1998,Menendez2008}. Observations of the M\textsubscript{s} vs. lattice expansion and M\textsubscript{s} vs. disorder relationships under mechanical stress-induced disordering processes have been reported \cite{Hernando1998,Menendez2008,Surinach1997,Gialanella1998}, however without consensus; the induced M\textsubscript{s} has been considered purely a disordering effect \cite{Zamora2009,Huffman1967,Takahashi1990}, and contradicted by claims of an M\textsubscript{s} contribution from the lattice expansion \cite{Nogues2006,Apinaniz2004}. 

Disorder caused by mechanical deformation tends to be concentrated at the strained regions, and can be spatially inhomogeneous and difficult to characterize. A more direct way to induce atomic rearrangements is \textit{via} ion-irradiation of thin films. Knock-on collisions with energetic ions can displace atoms from their ordered lattice sites, followed by a thermally-driven stochastic vacancy recombination leading to the formation of antisite defects. The mass of the penetrating ions, energy of the ions as well as temperature determine the chemical disorder manifested by the irradiation process – all of which can be exploited to subtly vary the induced disorder. This direct-disordering approach can be used to tailor the order-disorder transition in fine steps while keeping the composition fixed.

Here we show that the magnetic behavior of systematically disordered B2 Fe\textsubscript{60}Al\textsubscript{40} falls into three distinct regimes; despite the monotonic increase of $a_0$ with chemical disordering, the film behaves largely paramagnetic below a critical value of disordering, whereas above the critical regime it becomes ferromagnetic and M\textsubscript{s} is largely constant. The two regimes are separated by a third one, showing a critical M\textsubscript{s} increase.

Polycrystalline Fe\textsubscript{60}Al\textsubscript{40} films with a thickness of 250\,nm were deposited by single-target magnetron sputtering under a 3$\cdot10^{-3}$\,mbar Ar atmosphere on Si(001) buffered with 250\,nm thick SiO$_2$. The use of thin films allows the whole film volume to be chemically disordered by ions and subsequently probed by X-ray and magnetic measurements. Post-annealing at 773\,K for 1\,hr was performed  in vacuum to obtain B2 Fe\textsubscript{60}Al\textsubscript{40}. To achieve a systematic characterization of structure-property relationships, ion-irradiation of the above B2 ordered films was performed under a wide variety of conditions. The variable parameters were the ion-species (H\textsuperscript{+}, He\textsuperscript{+}, and Ne\textsuperscript{+}), ion-energy (17\,-\,170\,keV), ion-fluence (up to 4$\cdot10^{17}$\,ions/cm\textsuperscript{2}) and sample temperature during irradiation (100\,-\,523\,K). These parameters were selected based on Monte Carlo type simulations implementing the binary collision approximation \cite{Ziegler2010}, to achieve a peak average displacement between 0.07 and 5.77 per atom (for details see Supplement \cite{Supplement}).

\begin{figure}
  \hfill %
  \begin{flushleft}
  \centering
  \captionsetup[subfloat]{justification=raggedright,singlelinecheck=false,font=scriptsize,indent=0pt,margin=0pt,width=0.295\textwidth,labelformat=empty} 
  \subfloat[]{\includegraphics[width=8cm]{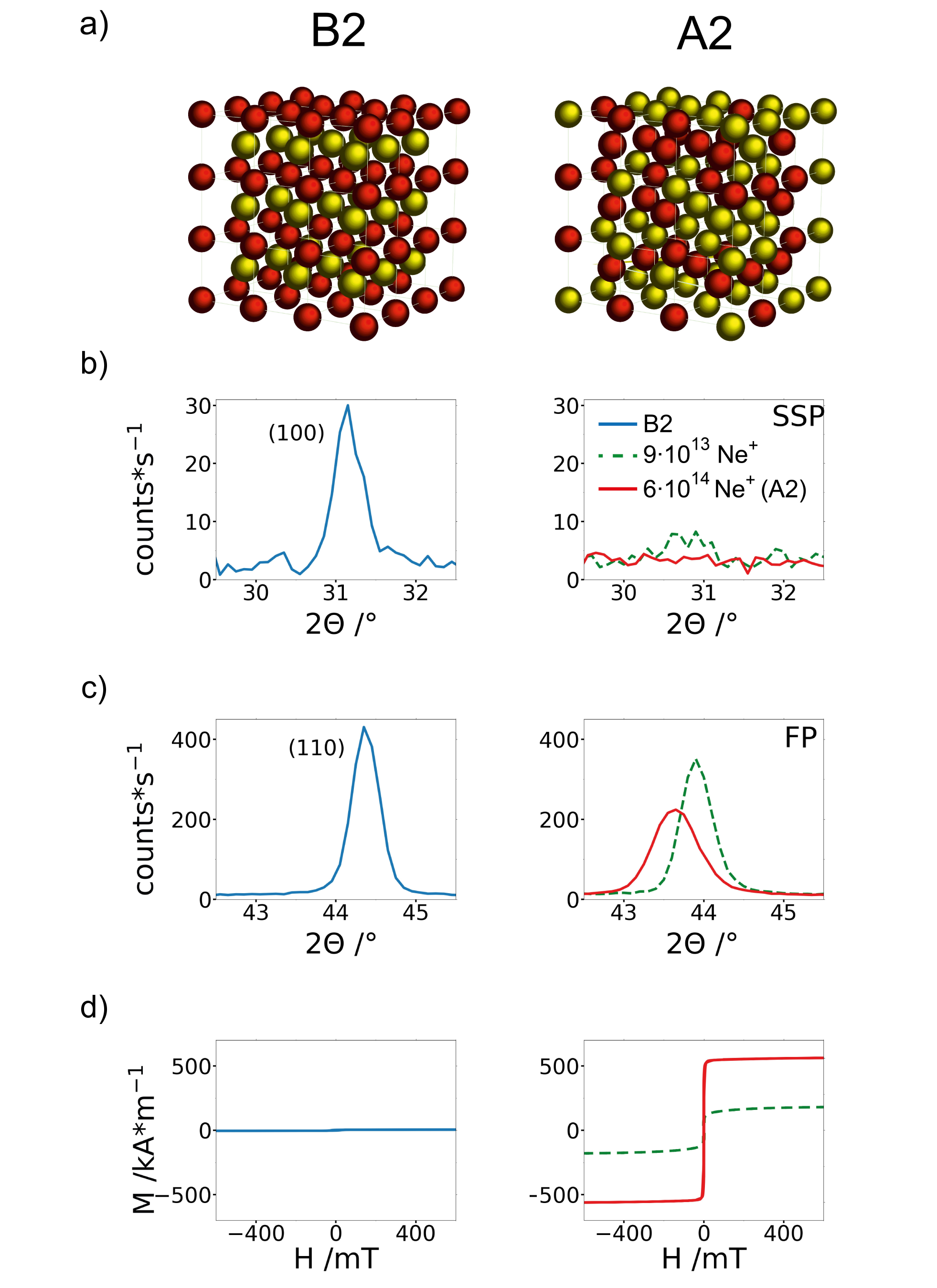}}
  \hfill
  \end{flushleft}
   \caption{\textbf{Disorder induced magnetic and structural changes. (a)} Schematic of the chemically ordered B2 structure (left) and the disordered A2 structure. \textbf{(b)} XRD measurement of the (100) superstructure peak (SSP) for various treatments. \textbf{(c)} Measured fundamental peak (FP). \textbf{(d)} Corresponding increase of the saturation magnetization (M\textsubscript{s}). The irradiations have been performed at 170\,keV.}
  \label{fig:1}
\end{figure}

Figures\,1\,b-d show the structural and magnetic analysis on selected samples with 3 different treatments – B2 and Ne\textsuperscript{+} irradiated with 9$\cdot10^{13}$ and 6$\cdot10^{14}$\,ions/cm\textsuperscript{2} leading to the fully ordered, intermediate, and disordered state, respectively. The lattice parameter ($a_0$) and order parameter $S$ were estimated using X-ray diffraction, where the shift of the (110) fundamental peak (FP) is a measure of the $a_0$, and the integral intensity of the (100) superstructure peak (SSP) is directly dependent on the ordering (see supplement). In Figure\,1, the XRD peaks of the fully ordered, intermediate and disordered thin films refer to order parameters of $S$=0.8, 0.4, and $<$0.2 (Figure\,1b), and lattice parameters of $a_0$=2.89\,\AA, 2.91\,\AA, and 2.93\,\AA\,(Figure\,1c), respectively. Correspondingly, the M\textsubscript{s} increases from 10\,kA/m for the fully ordered structure, to 180\,kA/m for intermediate order and $\sim$500\,kA/m for the fully disordered (Figure\,1d). All measurements have been performed at room temperature. 

The order parameter $S$ can be estimated using the square root of the ratio of the integrated measured intensities of the SSP, $I_{SSP}$ and of the FP, $I_{FP}$ with respect to the theoretical calculated values of the ordered B2 Fe\textsubscript{60}Al\textsubscript{40} structure \cite{Suryanarayana1998}:

\begin{equation}
S=\sqrt{\frac{({I_{SSP}}/{I_{FP}})_{measured}}{({I_{SSP}}/{I_{FP}})_{calculated}}}
\label{eq:S}
\end{equation}

where S is dimensionless and 0 for a fully disordered A2 structure, and due to the off-equiatomic composition, can reach only a maximum of $\sim$0.8 for the best ordered case \cite{Warren1990}. Since ion-irradiation gradually disorders the film, a more convenient term is the disorder parameter, defined as \hbox{1-$S$}, and will be used in the discussion. The order parameter was estimated from the low order Bragg reflections. Considering the background of the XRD measurement and the peak broadening due to variations of micro\-strain and crystallite size, an estimated disorder of up to \hbox{1-$S$}\,$\approx$\,0.8 is detectable (Supplement). 

\begin{figure*}
  \hfill %
  \begin{flushleft}
  \centering
  \captionsetup[subfloat]{justification=raggedright,singlelinecheck=false,font=scriptsize,indent=0pt,margin=0pt,width=0.295\textwidth,labelformat=empty} 
  \subfloat[]{\includegraphics[width=16cm]{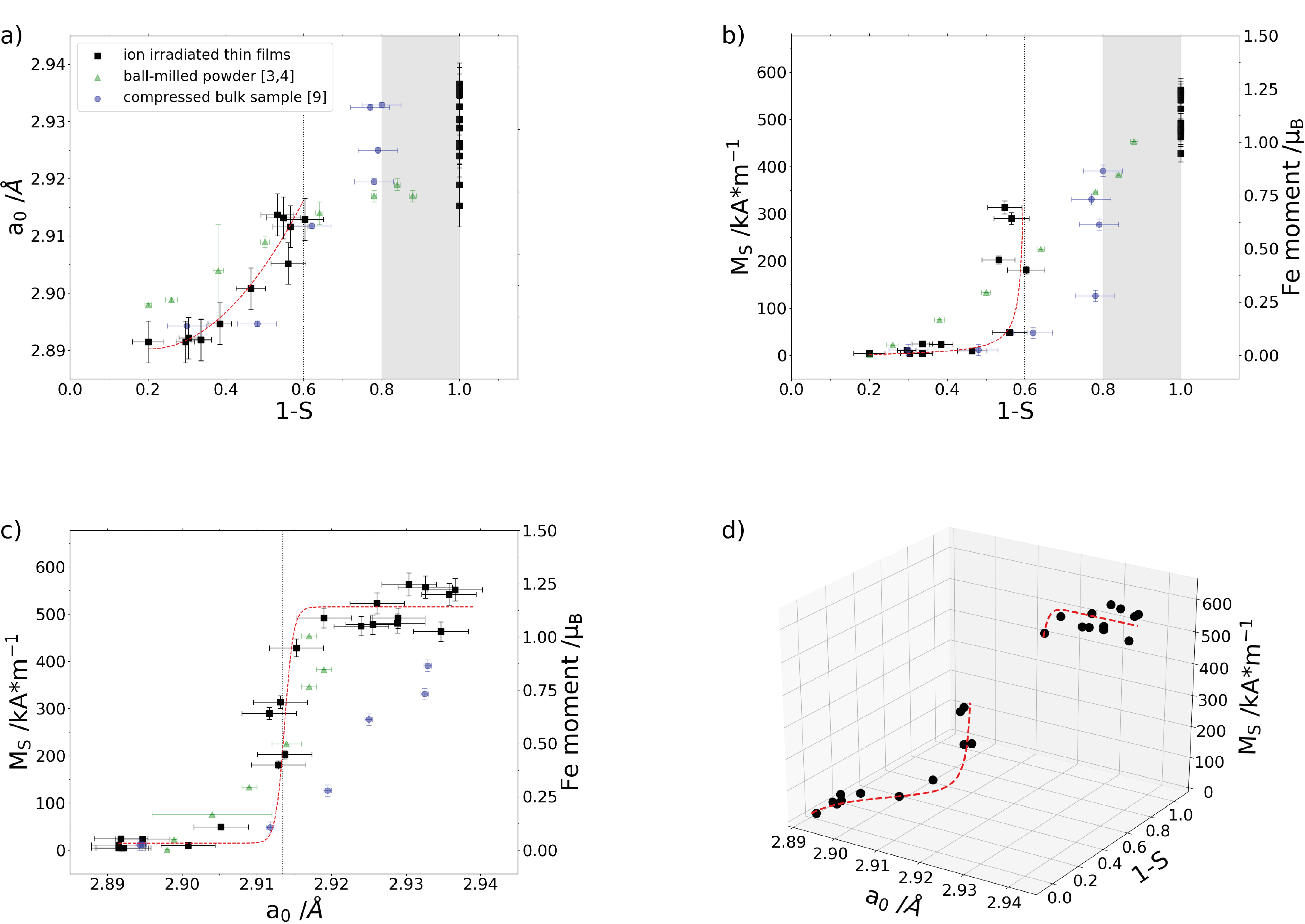}}
  \hfill %
  \end{flushleft}
  \caption{\textbf{Magnetic and structural properties in disordered Fe\textsubscript{60}Al\textsubscript{40}. (a)} Lattice parameter $a_0$ as a function of the disorder parameter \hbox{1-$S$}. \textbf{(b)} Dependence of the saturation magnetization M\textsubscript{s} on the disorder. \textbf{(c)} Variation of the magnetization with lattice parameter. \textbf{(d)} 3-dimensional plot showing the inter-dependence of the studied magnetostructural properties on thin films. (Black squares) obtained results for ion-irradiated thin films. The red dashed lines serve as a guide to the eye and the black dotted lines mark the critical point. (Green triangles) and (Blue circles) illustrate the published interrelationships for ball-milled powder \cite{Hernando1998,Amils1998} and a compressed bulk sample \cite{Menendez2008}, respectively. The range for 1-$S$\,$>$\,0.8 is shaded grey indicating the nominally fully disordered films.}  %
  \label{fig:2} %
\end{figure*}

We evaluate the inter-dependencies of the structural and magnetic properties, namely, \hbox{1-$S$}, $a_0$ and M\textsubscript{s}. The $a_0$ and M\textsubscript{s} are plotted as functions of \hbox{1-$S$} in Figure\,\ref{fig:2}a and b respectively, whereas the M\textsubscript{s}($a_0$) is shown in Figure\,2c. Despite the vast variety of conditions applied in the experiments, the relationship between M\textsubscript{s}, $a_0$ and \hbox{1-$S$} collapses into a single curve, as shown in Figure\,\ref{fig:2}d.

As seen in Figure\,\ref{fig:2}a, $a_0$ increases monotonically with disorder, from $a_0$\,=\,2.89\,\AA\,for the fully ordered films to 2.91\,\AA\,for \hbox{1-$S$}\,=\,0.6. Further ion-irradiation results in a vanishing SSP, implying a \hbox{1-$S$}\,$>$\,0.8. Intermediate values between \hbox{1-$S$}\,=\,0.6 and 0.8 were not observed under any of the attempted experimental conditions. Thin films with \hbox{1-$S$}\,$>$\,0.8 are considered nominally fully disordered. Here, $a_0$ varies over a range from 2.91\,\AA\,to 2.935\,\AA. 

The M\textsubscript{s} for fully ordered and thus largely paramagnetic films is initially constant (10 kA/m) with increasing disorder (Figure 2b), until a sharp onset close to \hbox{1-$S$}\,=\,0.6 appears, after which a sharp increase of M\textsubscript{s} to 300 kA/m is observed. Above this value only the fully disordered structures could be observed by XRD. M\textsubscript{s} shows a similar dependence on $a_0$ (Figure 2c): initially, $a_0$ increases from 2.89 to 2.91 A, whereas M\textsubscript{s} remains approximately constant. A further increase in $a_0$ leads again to a step-like increase of M\textsubscript{s} up to its maximum. With further increasing $a_0$, M\textsubscript{s} remains around its maximum value.

The two regimes of different, but nearly stable M\textsubscript{s} are demarcated by a dotted line passing through the critical values of \hbox{1-$S$}\,=\,0.6 and $a_0$\,=\,2.91 \AA\,(Figure\,2). Even as M\textsubscript{s} and \hbox{1-$S$} reach their limits above the critical point, the $a_0$ can be further increased until reaching its maximum; this can be explained by the further disordering of residual short-range ordered B2 regions since the irradiation-induced lattice expansion due to vacancies can be neglected \cite{Liedke2015}.

We compare our results with previous approaches using different methods of mechanical stress-induced disordering of bulk-like B2 Fe\textsubscript{60}Al\textsubscript{40}, \textit{i.e.} ball-milled powder \cite{Hernando1998,Amils1998} and almost uniaxial compressed bulk samples \cite{Menendez2008}. As seen in Figure\,2b, below and above \hbox{1-$S$}\,$=$\,0.6 there is agreement between the mechanical stress-induced and direct-disordering approaches. However, at \hbox{1-$S$}\,$=$\,0.6 no critical behavior is observed for the mechanical stress-induced approaches. The possible inhomogeneities within the sample volume of the mechanical stress-induced disordered material, especially for ball-milling, may result in a smoothed 1-$S$ behavior.

In general, intermediate states of disorder ranging from \hbox{1-$S$}\,$=$\,0.6 to fully disordered (\hbox{1-$S$}\,$=$\,0.8 to 1) have not been observed. This is true for mechanically disordered bulk samples in literature as well as the present irradiation-disordered films. The present investigation of the region around the critical point reveals that whereas the monotonic behavior of $a_0$ vs. \hbox{1-$S$}, the M\textsubscript{s} vs. \hbox{1-$S$} shows an unambiguous critical increase (Figure\,2b). 

Having established this critical behavior for the onset of ferromagnetism, we examine the plausibility of this behavior using density functional theory (DFT). Several calculations are mentioned in literature \cite{Gu1992,Kulikov1999,Apinaniz2001,Apinaniz2003,Apinaniz2004,Chacham1987,Bose1997,Reddy2000}, with each approach partially explaining the observed behavior.

Kulikov \textit{et al.} \cite{Kulikov1999} applied tight-binding linear muffin-tin orbital (TB LMTO) approach on B2 Fe\textsubscript{50}Al\textsubscript{50} and obtained a moment ($\mu_{Fe}$) of 0.76\,$\mu_{B}$ on the Fe atom, and an equilibrium $a_0$\,$\approx$\,2.86\,\AA. The calculation also yielded a linear increase of $\mu_{Fe}$ with increasing $a_0$. The calculation however, does not reproduce the increasing $a_0$ with disorder, seen clearly in Figure\,2a, as well as in other works \cite{Hernando1998,Amils1998,Menendez2008,Roeder2015,Smirnov2005,Bose1997,Reddy2000}. An increase in the number of Fe-Fe nearest neighbors at the antisite causes an increase in the occupancy of the $d$\,band. The increase of the spin-polarization at the disorder site due to electron filling is known from Kulikov \textit{et al.'s} rigid band picture \cite{Kulikov1999}. The perturbation at the antisite is associated with an increased $a_0$ as well as Friedel oscillations that cause a further increase of $\mu_{Fe}$ of Fe atoms that are a few atomic spacings away from the antisite. The rigid band picture is consistent with the monotonic variation of $a_0$ and \hbox{1-$S$}, while $\mu_{Fe}$ remains at minimum (Figure\,2a). Below the critical point, the system behaves paramagnetic as it would still be B2 ordered, seen experimentally in Figure\,2c.

The regime observed above the critical point cannot be explained by the rigid band picture. Here the effect of the lattice expansion on the DOS must be considered. Apiñaniz \textit{et al.} \cite{Apinaniz2001,Apinaniz2003,Apinaniz2004} applied TB LMTO method to both B2 and A2 structures and showed an increased $a_0$ with disordering; the calculated equilibrium $a_0$ for the B2 and A2 are 2.84 and 2.89\,\AA\,respectively, and $\mu_{Fe}$ of 0.64 and 1.7\,$\mu_{B}$ respectively. Furthermore, a critical behavior of the $\mu_{Fe}$ with increasing $a_0$ is predicted, whereby $\mu_{Fe}$ in B2 Fe\textsubscript{50}Al\textsubscript{50} rises sharply from zero to 0.5\,$\mu_{B}$ as the $a_0$ expands above 2.78\,\AA; in the absence of disorder. Whereas the calculated critical dependence of $\mu_{Fe}$ on $a_0$ is inconsistent with the results of mechanical stress-induced processes in the literature, it does bear resemblance to the observations on ion-irradiated films shown here.

The prediction that the B2 structure can undergo a transition to a ferromagnetic state above a critical $a_0$, even without disorder, can prove useful in explaining the current experimental observations. 

We explore the above aspect by first performing DFT calculations on the relevant composition, \textit{i.e.} B2 Fe\textsubscript{60}Al\textsubscript{40}. First principles density functional theory (DFT) calculations using fully relativistic Korringa-Kohn-Rostoker (KKR) formalism with the SPRKKR package \cite{Ebert2011} were performed. Perdew-Burke-Ernzerhof (PBE) exchange-correlation functional has been used within generalized gradient approximation. Configurational disorder was treated within coherent potential approximation (CPA).

The effect of increasing $a_0$\ on the Fe moment for B2 as well as A2 structures is shown in Figure\,3a. Whereas the A2 structure is FM throughout the investigated $a_0$\ range, lattice expansion causes a ferromagnetic onset in the B2 at $a_0$\ $\approx$\,2.87\,\AA. According to calculations, the equilibrium $a_0$ for the B2 structure lies close to 2.87\,\AA, with the $E_F$ is located within a narrow pseudogap of $\approx$\ 1 eV width, thereby rendering the spin-splitting highly sensitive to  changes in the DOS that can be manifested by the increasing disorder and  $a_0$. Figures\,3b and c show the effect of disorder and lattice expansion respectively on the DOS.

\begin{figure}
  \hfill %
  \begin{flushleft}
  \centering
  \captionsetup[subfloat]{justification=raggedright,singlelinecheck=false,font=scriptsize,indent=0pt,margin=0pt,width=0.295\textwidth,labelformat=empty} 
  \subfloat[]{\includegraphics[width=8.2cm]{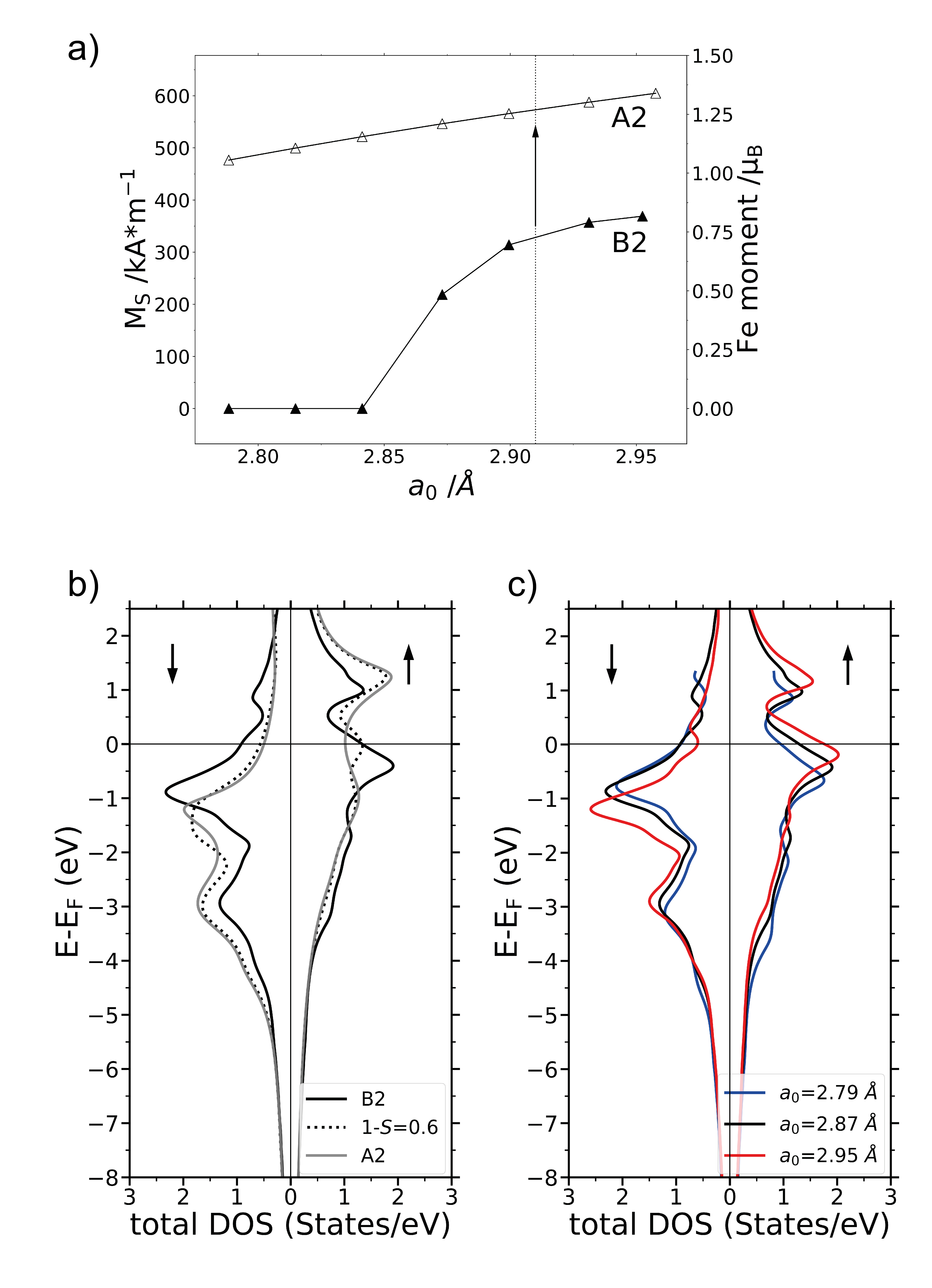}}
  \hfill
  \end{flushleft}
\caption{\textbf{DFT calculation results for Fe\textsubscript{60}Al\textsubscript{40}. (a)} Variation of the saturation magnetization or magnetic moment per Fe atom with the lattice parameter for the B2 (filled triangles) and A2 (empty triangles) phase.The dotted line indicates the experimentally observed critical $a_0$, where the  transition from para- to ferromagnetic behavior occurs. \textbf{(b)} DOS of the B2-ordered (black line), partially disordered,\hbox{1-$S$} = 0.6 (dotted line) and fully-disordered, A2 (grey line) structures.\textbf{(c)} Variation in DOS of the disorder-free B2 structure, due to lattice expansion.} %
  \label{fig:3} %
\end{figure} 
 
Disorder causes a smearing of the DOS which is expected due to scattering from antisites. As seen in Figure\,3b, the consequent changes to the DOS at the  $E_F$ are sufficient to cause a spin-splitting, where the partially disordered state of \hbox{1-$S$} = 0.6 shows an Fe-moment of 1.25\,$\mu_{B}$. Comparing the DOS for the partially disordered state to that of the fully disordered A2 structure, it is seen that the spin-splitting due to antisite-scattering saturates at \hbox{1-$S$} = 0.6. This matches well with the experimentally observed critical point above which the M\textsubscript{s} becomes independent of the disordering (Figure\,2b).  

Similarly, Figure\,3c considers the effect of lattice expansion on disorder-free B2 Fe\textsubscript{60}Al\textsubscript{40}. Below the critical $a_0$, the close distances between the atoms can cause a smearing of the $d$\,bands. As the lattice expands, orbital hybridization reduces and the peaks in the $d$\,band start to narrow. The location of $E_F$ in the vicinity of the narrowing $d$\,band peaks can, at a critical point and above, make spin-splitting energetically favorable. The onset of ferromagnetism in disorder free B2 Fe\textsubscript{60}Al\textsubscript{40} is therefore due to the particular position of $E_F$ in the presence of narrowing $d$\,band peaks. Since the $\mu_{Fe}$ increases with lattice expansion in both the B2 as well as the A2 structures (Figure\,3a), the increasing $\mu_{Fe}$ caused by band narrowing appears to be valid for any given state of disorder.

From the above DOS considerations, it is seen that both antisite-scattering and band-narrowing favour an increased spin-splitting. The contribution of antisite-scattering to the spin-splitting saturates at \hbox{1-$S$} = 0.6, whereas increasing $a_0$ tends to continuously increase the spin-splitting, both in the fully disordered as well as the residual B2 ordered regions \cite{Torre2018}. The initial sparse disordering of the B2 lattice leads to localized $\mu_{Fe}$ at antisites, manifesting an interplay with the disorder and $a_0$. The strain induced due to the lattice expansion of the disordered regions increases the average $a_0$ thus modifying the DOS and causing spin-splitting throughout the lattice. The M\textsubscript{s} will follow a path bound by the B2 and A2 lines, indicated by the arrow in Figure\,3a.

The latter part of the transition where M\textsubscript{s} is solely dependent on the $a_0$ has been addressed in previous studies, arriving at a conclusion that the lattice expansion contributes about 35\,\% of the induced $\mu_{Fe}$ \cite{Nogues2006}. However, as we have seen in the above discussion, separating the respective contributions of disorder and lattice expansion, is valid only in the regime above the critical point. 

Unraveling the interplay between the disorder induced moment and lattice expansion, as well as the critical behavior sheds light on the magnetism of disordered systems, and can be applicable to a broad range of binary alloys. Our results show that controlled disordering of alloys can be a promising approach to sensitively engineer the DOS of alloys and achieve tailored functional properties.

\nocite{Moeller2017}

\vspace{1cm}
We acknowledge the assistance of Andrea Scholz for structural analysis. We thank Johannes von Borany for useful discussions. Irradiation experiments were performed at the Ion Beam Center of the Helmholtz-Zentrum Dresden - Rossendorf. Funding from DFG Grants BA 5656/1-1 \& WE 2623/14-1 is acknowledged. B.S. acknowledges financial support from Swedish Research Council and Swedish National Infrastructure for Computing for allocation of computing time under the project SNIC2017-12-53.

\end{document}